\documentstyle[12pt]{article}

\begin{document}

\begin{titlepage}
\vspace{1cm}

\begin{centering}

{\Large \bf On the BFFT quantization of first order systems}

\vspace{.5cm}
\vspace{1cm}

Ricardo Amorim    \\
Instituto de F\'{\i}sica\\ 
Universidade Federal do Rio de Janeiro\\ 
RJ 21945-970 - Caixa Postal 68528 - Brasil\\
Ronaldo Thibes     \\
Instituto de F\'{\i}sica\\
Universidade Federal do Rio de Janeiro\\ 
RJ 21945-970 - Caixa Postal 68528 - Brasil\\
and\\
Instituto de Ci\^encias Exatas e da Natureza\\
Universidade do Grande Rio Professor Jos\'e de Souza Herdy\\
Rua Prof. Jos\'e de Souza Herdy, 1160\\
25 de Agosto - Duque de Caxias - RJ 25073-200 - Brasil\\

\begin{abstract}
By using the field-antifield formalism, we show that the method of
Batalin, Fradkin, Fradkina and Tyutin to convert Hamiltonian systems
submitted to second class constraints introduces compensating fields 
which do not belong to the BRST cohomology at ghost number one. This assures
that the gauge symmetries which arise from the BFFT procedure are not obstructed at quantum level. An example where massive electrodynamics is coupled to chiral fermions is considered. We solve the quantum master equation for the model and show that the respective counterterm has a decisive role in extracting anomalous expectation values associated with the divergence of the Noether chiral current. 
\end{abstract}
\end{centering}
\vfill

\noindent amorim@if.ufrj.br, thibes@if.ufrj.br
\vspace{1cm}
\end{titlepage}
\pagebreak

\section{Introduction}
The seminal works of Dirac \cite{Dirac} on constraint Hamiltonian systems
have been developed in several important research lines. One of these 
developments is due to Fradkin, Fradkina, Batalin and Tyutin
(BFFT) \cite{BFFT}, where Hamiltonian systems submitted to second class constraints are conveniently considered. The method of BFFT consists in
enlarging the original phase-space of the theory by adding compensating fields which permit to convert the second class constraints into first class
ones. In doing so, it is possible to avoid Dirac brackets  which
can present severe  problems when one follows the canonical approach to quantization \cite{HT}. As first class constraints are also necessarily associated with local gauge symmetries, a system converted by the BFFT procedure can
be treated by using all the machinery associated with the BRST formalism
\cite{BRST}.
 
The BRST approach for quantization of gauge theories appears with all its
power in the field-antifield formalism \cite{BV,HT,GPS}. This formalism gives an elegant and
systematic way for constructing the functional generator of any general gauge theory, with possible reducible or open gauge algebras. At the same time, eventual obstructions to the gauge symmetries due to quantum
effects are naturally taken in account inside the field-antifield formalism.
\bigskip

In this work we consider, by using  some tools of the field-antifield formalism, the quantization of first order gauge theories which have been obtained from
second class constrained systems by the process of
conversion developed by BFFT. We show that the compensating fields introduced by the conversion procedure do not belong to the BRST cohomology
\cite{HT} at ghost number one. So  there is no possible term
in the space of fields and antifields with ghost number one and BRST closed not being BRST exact. This means that the Wess-Zumino consistence condition \cite{GPS} is solved in a trivial way: there is no gauge anomaly for such class of systems and the quantum master equation can always be solved
with the inclusion of a proper counterterm in the quantum action.
It is useful to observe that this counterterm, if it exists, can play a non-trivial role. We give an example where massive electrodynamics couples to chiral fermions. There we show that it is necessary to introduce  a non-trivial counterterm in order to solve the quantum master equation. This counterterm permit us to extract an anomalous expectation value related to the divergence of the fermion Noether chiral current.

We would like to note that
compensating fields have been largely employed directly inside  Lagrangian
descriptions \cite{Compensating,dWG}. There the purpose is not converting second class constraints, but  to enlarge the symmetry content of a theory in such a way that the original description is recovered within some gauge choice. Under this last point of view, BFFT and Lagrangian compensating fields play similar roles. In several examples of Lagrangian descriptions it is proved that compensating fields also do not belong to the cohomology
at ghost number one and can be used as well to extract anomalous expectation values
of physically relevant quantities \cite{us}.

We organized this work as follows: In section {\bf2} we present a brief review of the BFFT conversion of first order systems submitted to pure
second class constraints. We display the local gauge invariance of the
first order action which is introduced by the BFFT compensating fields.
The functional quantization of such a system is described in section {\bf3}, 
by using the tools of the field-antifield formalism. We derive the BRST differential and explicitly show that the BFFT variables do not belong to
the BRST cohomology at ghost number one. This assures that the quantum master equation can be solved for any system of this class. In section {\bf4} the ideas presented in the first sections are applied to a model which
describes massive electrodynamics coupled to chiral fermions in four space-time dimensions. By using a regularization that keeps the vector symmetry as a preferential one, the quantum master equation is solved with the introduction of an specific counterterm in the quantum action. A few different gauge fixing choices are explored and covariant actions are obtained. When the gauge freedom is fixed by identifying the compensating fields with external functions, we show that the independence of the
path integral with respect to those external functions permit us to derive
expectation values which are related to the anomalous divergence of the Noether chiral current.
We reserve section {\bf5} to some general comments and concluding remarks.

\bigskip

\section{First order systems submitted to second class constraints}
\setcounter{equation}{0}
\bigskip 

In this section we will review a few topics on constrained Hamiltonian 
systems \cite{HT} and on the BFFT conversion procedure  \cite{BFFT} in order to fix 
notations and to introduce some results that will be useful for further developments.
Let us start by considering a generic first order system living in a (phase) space with
discrete bosonic  coordinates 
$y^\mu$, $\mu=1,2,...,2N$. The extension to more general situations can be trivially done.
Its action is written as 

\begin{equation}
\label{z1}
S_0=\int\,dt\,\left(B_\mu\dot y^\mu-\lambda^\alpha\chi_\alpha-H\right)\,
\end{equation}

\noindent where $B_\mu,\,H$ and $\chi_\alpha$ are in principle arbitrary functions of the coordinates but do not depend on the velocities. The Lagrange multipliers $\lambda^\alpha$ are to be regarded as independent quantities.
From the above expression one can read the symplectic form

\begin{equation}
f_{\mu\nu}={{\partial B_\nu}\over{\partial y^\mu}}-{{\partial B_\mu}\over{\partial y^\nu}}
\label{z2}
\end{equation}

\noindent which has an inverse  $f^{\mu\nu}$ if the system is well defined. With its aid, we can define the  brackets between any
two functions $A(y)$ and $B(y)$ as

\begin{equation}
\bigl\{A,\,B\bigr\}={\partial A\over{\partial y^\mu}}
f^{\mu\nu}{\partial B\over{\partial y^\nu}}\label{z3}
\end{equation}
It follows that

\begin{equation}
\bigl\{y^\mu,\,y^\nu\bigr\}=f^{\mu\nu}\label{z4}
\end{equation}

The brackets appearing in the above expressions can be interpreted as Poisson brackets only in a broad sense, since  they take in
account the primary second class constraints of the Dirac's scheme \cite{FJ}. In this sense they are primary Dirac brackets.
Let now $H$  and  $\chi_\alpha$,
$\alpha=1,2,...,2n$, represent respectively a first class
Hamiltonian  and a set of second class constraints.
The Hamiltonian and the constraints then satisfy 
the structure 

\begin{eqnarray}
\bigl\{\chi_\alpha,\chi_\beta\bigr\}&=&\Delta_{\alpha\beta}
\nonumber\\
\bigl\{H,\chi_\alpha\bigr\}&=&
V_\alpha^\beta\chi_\beta\label{3}
\end{eqnarray}

\bigskip  \noindent As the $\chi$'s are  second class, 
the constraint
matrix $\Delta_{\alpha\beta}$ is regular. 
\bigskip

It may be convenient to extend the phase-space by adding  compensating variables $\phi^\alpha$, $\alpha=1,2,\dots,2n$, but at the same time converting the set of second class  constraints into a first-class one. This assures that the number of degrees of freedom is not changed by the process, which also introduces local symmetries that permit one to quantize
the theory by using the powerful tools of local gauge  theories. 

To perform this conversion through the
BFFT procedure, it is assumed that the BFFT compensating variables $\phi^\alpha$
satisfy fundamental brackets given by

\begin{equation}
\bigl\{\phi^\alpha,\phi^\beta\bigr\}=\omega^{\alpha\beta}
\label{6}
\end{equation}

\bigskip
\noindent where $\omega$ is 
some constant, antissymmetric and invertible  matrix. 
In order to avoid the introduction of further second class constraints, it may be convenient to
choose $\omega$ in such a way that the compensating variables form a set of canonical conjugated quantities.
In any case, it follows that in the BFFT extended space, the brackets between any two quantities
$A(y,\phi)$ and $B(y,\phi)$ are written as

\begin{equation}
\bigl\{A,\,B\bigr\}={\partial A\over{\partial y^\mu}}
f^{\mu\nu}{\partial B\over{\partial y^\nu}}+{\partial A\over{\partial \phi^\alpha}}
\omega^{\alpha\beta}{\partial B\over{\partial \phi^\beta}}
\label{7}
\end{equation}

\bigskip
\noindent as both sectors  are
 independent.
\bigskip

The general idea of the BFFT algorithm is to replace the old set of second class constraints and the old Hamiltonian by a new set of first class constraints
$\tilde\chi_\alpha = \tilde\chi_\alpha(y,\phi)$ and Hamiltonian 
$\tilde H=\tilde H(y,\phi)$
in such a way that they become involutive:

\begin{eqnarray}
\bigl\{\tilde\chi_\alpha,\tilde\chi_\beta\bigr\}&=&0
\nonumber\\
\bigl\{\tilde H,\tilde\chi_\alpha\bigr\}&=&0
\label{8}
\end{eqnarray}

\bigskip
By requiring that  $\tilde A(y,0)=A(y)$ for any quantity $A$ defined  in the extended space,  
it is assured that the original formulation of the  theory is recovered 
when the unitary
gauge $\phi^\alpha=0$ is implemented. 
In Refs. \cite{BFFT} it is proved that Eqs. (\ref{8}), submitted to the
above condition, always have a power series solution in the compensating variables,
with coefficients with only $y^\mu$ dependence. 
The second class constraints, for instance, can be extended to

\begin{equation}
\tilde\chi_\alpha(y,\phi)=\chi_\alpha(y)+X_{\alpha\beta}(y)\phi^\beta
+X_{\alpha\beta\gamma}(y)\phi^\beta\phi^\gamma+\dots
\label{9} 
\end{equation}

\bigskip
Conditions (\ref{8}) impose restrictions
on the expansion coefficients. As an  example, 
the regular matrices $X_{\alpha\beta}$ must satisfy the identity

\begin{equation}
\label{10}
X_{\alpha\beta}\omega^{\beta\gamma}X_{\delta\gamma}=-\Delta_{\alpha\delta}
\end{equation}

\bigskip
Even if some quantity $A(y)$ is not a second class constraint,  it
can also be extended to $\tilde A(y,\phi)$ in order to be involutive with the converted constraints $\tilde\chi_\alpha$. Following the
BFFT procedure we can show that in this situation

\begin{equation}
\tilde A(y,\phi)=A(y)- \phi^\alpha\omega_{\alpha\beta}X^{\beta\gamma}\{\chi_\gamma,A\}+...
\label{10a}
\end{equation}

\bigskip
\noindent where the dots represent al least second order corrections in $\phi$ to $A(y)$. In (\ref{10a}), the matrix $X$ with contravariant indices is to be considered as the inverse of the corresponding covariant one.  
Now it is possible to prove that the first order action

\begin{equation}
S_0=\int\,dt\,[B_\mu \dot y^\mu+B_\alpha\dot\phi^\alpha-\lambda^\alpha\tilde\chi_\alpha-\tilde H]
\label{11}
\end{equation}

\bigskip
\noindent is invariant under the gauge transformations

\begin{eqnarray}
\delta y^\mu&=&\{y^\mu,\tilde\chi_\alpha\}\epsilon^\alpha\nonumber\\
\delta\phi^\alpha&=&\{\phi^\alpha,\tilde\chi_\beta\}\epsilon^\beta
\nonumber\\
\delta\lambda^\alpha&=&\dot\epsilon^\alpha
\label{12}
\end{eqnarray}

\noindent By using the Jacobi Identity and Eqs. (\ref{8}) we see that (\ref{12}) close in an Abelian
algebra. As in (\ref{z2}), in (\ref{11}) $B_\alpha$
is related to the inverse of $\omega^{\alpha\beta}$ through

\begin{equation}
\omega_{\alpha\beta}={{\partial B_\beta}\over{\partial\phi^\alpha}}- 
{{\partial B_\alpha}\over{\partial\phi^\beta}}
\label{13}
\end{equation}
\bigskip

\noindent One can always choose 
$B_\alpha={1\over2}\omega_{\alpha\beta}\phi^\beta$ 
without loss of generality.
By using some of the above equations, it is not difficult to show that

\begin{equation}
\delta[B_\mu \dot y^\mu+B_\alpha\dot\phi^\alpha-\lambda^\alpha\tilde\chi_\alpha-\tilde H]
={d\over{dt}}\{[B_\mu f^{\mu\nu}{{\partial\tilde\chi_\alpha}\over{\partial y^\nu}}+
B_\beta \omega^{\beta\rho}{{\partial\tilde\chi_\alpha}\over{\partial \phi^\rho}}
-\tilde\chi_\alpha]\epsilon^\alpha\}
\label{14}
\end{equation}

\bigskip
\noindent and consequently  (\ref{11}) is indeed invariant under the local gauge transformations (\ref{12}), provided boundary terms can be discarded. 
\bigskip

\section{Quantization}
\setcounter{equation}{0}
\bigskip

Let us perform the quantization of the system described above  along the field-antifield formalism \cite{BV,HT,GPS}. To do so it is first necessary to introduce  antifields $\Phi^*_A=(y^*_\mu,\phi^*_\alpha,\lambda^*_\alpha,c^*_\alpha)$ corresponding to the fields
$\Phi^A=(y^\mu, \phi^\alpha, \lambda^\alpha,c^\alpha)$. In our case, 
$y^\mu$, $\phi^\alpha$ and $\lambda^\alpha$ are bosonic and have ghost number zero. The ghosts $c^\alpha$ are fermionic and have ghost number one. The corresponding antifields have opposite grassmanian parity and ghost number given by minus the ghost number of the corresponding field minus one. One can verify that the field-antifield action

\begin{equation}
\label{15}
S= S_0+\int\,dt\,[y^*_\mu\{y^\mu,\tilde\chi_\alpha\}c^\alpha+
\phi^*_\beta\{\phi^\beta,\tilde\chi_\alpha\}c^\alpha+\lambda^*_\alpha\dot c^\alpha]
\end{equation}

\bigskip
\noindent satisfies then the classical master equation

\begin{equation}
\label{16}
{1\over 2}(S,S)=0
\end{equation}

\bigskip\noindent 
where the antibracket between any two quantities $X[\Phi,\Phi^*]$ and $Y[\Phi,\Phi^*]$ is defined as
 $(X,Y) = {\delta_rX\over
\delta\Phi^A} {\delta_lY\over\delta\Phi^\ast_A}
- {\delta_rX\over \delta\Phi^\ast_A}
  {\delta_lY\over \delta\Phi^A}$. When pertinent, we are assuming the de Witt's notation of sum and integration over intermediary variables.

\bigskip
In the BV formalism, the BRST differential is introduced through

\begin{equation}
\label{a17}
s\,X=(X,S)
\end{equation}

\noindent for any local  functional $X=X[\Phi,\Phi^*]$. As a consequence of the master equation (\ref{16}) and Jacobi identity, $s$ is nilpotent. So, saying that the BV action satisfies the master equation is equivalent to say that it is BRST invariant. 

To fix a gauge we need to
introduce trivial pairs $\bar c_\alpha\,,\bar\pi_\alpha$ as new fields,
and the corresponding antifields ${\bar c}^{*\alpha},{\bar\pi}^{*\alpha}$,
as well as a gauge-fixing fermion $\Psi$. The antifields are eliminated by choosing  $\Phi^*_A = {{\partial\Psi}\over{\partial\Phi^A}}$. It is always possible to 
choose

\begin{equation}
\label{17}
\Psi=\bar c_\alpha\phi^\alpha
\end{equation}

\bigskip
\noindent associated with the unitary gauge, but different
choices can be done.
It is also necessary to extend the field-antifield action to a non-minimal one

\begin{equation}
\label{18}
S\rightarrow S_{nm}=S+\int\,dt\,\bar\pi_\alpha{\bar c}^{*\alpha}
\end{equation}

\bigskip
\noindent in order to implement the gauge fixing introduced by $\Psi$. The gauge-fixed vacuum functional is then  defined as

\begin{equation}
Z=\int[d\Phi^A][\det \omega]^{-{1\over2}} [\det f]^{-{1\over2}}
   \exp\{\frac{i}{\hbar}\,S_{nm}[\Phi^A,\Phi^*_A - {{\partial\Psi}\over{\partial\Phi^A}}]\}
\label{19}
\end{equation}

\bigskip
In the unitary gauge, we observe that
besides the identification $\bar c^{*\alpha}=\phi^\alpha,\,\phi^*_\alpha=
\bar c_\alpha$, all the other antifields vanish. With this and the use of
Eqs. (\ref{9}-\ref{10}), we see that formally (\ref{19}) reduces to 
the Senjanovic \cite{Senj} path integral

\begin{eqnarray}
Z&=&\int[dy^\mu]\vert \det\,f\vert^{-{1\over2}}\delta[\chi_\alpha]
\vert\det\Delta\vert^{1\over 2}\nonumber\\
& &\exp\left\{\frac{i}{\hbar}\int dt[B_\mu \dot y^\mu-
H]\right\}
\label{4}
\end{eqnarray}

\noindent Actually this reduction can only be done if
quantum effects do not  obstruct the
gauge symmetries. Possible obstructions are related to
the dependence of the path integral with respect to redefinitions of the gauge-fixing fermion $\Psi$. In general, if the classical
field-antifield action $S$ can be replaced by some quantum action $W$
expressed as a local functional of fields and antifields
and satisfying the so-called quantum master equation

\begin{equation}
\label{20}
{1\over 2}(W,W)\, - \, i\hbar\Delta W\
\,=\,0
\end{equation}

\bigskip
\noindent then the gauge symmetries are not obstructed at quantum level. 
In  expression (\ref{20}) we have introduced the potentially singular operator
$\Delta \equiv
{\delta_r\over\delta\Phi^A}{\delta_l\over\delta\Phi^\ast_A}$
and it was assumed that $W$ can be expanded in powers
of $\hbar$ as

\begin{equation} 
W[\Phi^A,\Phi^{\ast}_A ] = 
S[\Phi^A ,\Phi^{\ast}_A ] +
\sum_{p=1}^\infty \hbar^p M_p [\Phi^A ,\Phi^{\ast}_A ]
\end{equation}

\noindent  The two first terms of the quantum master equation (\ref{20}) are

\begin{eqnarray}
\label{21}(S,S) &=& 0\\
\label{22}
(M_1,S) &=& \,i\, \Delta S
\end{eqnarray}

As expected, the tree approximation gives (\ref{16}). Eq. (\ref{22}) is only formal, since the action of the operator
$\Delta$ must be regularized. If it vanishes when applied on $S$, the quantum action $W$ can be identified with $S$. If $\Delta S$ gives a non-trivial result but there exists
some $M_1$ expressed in terms of local fields such that (\ref{22}) is satisfied, gauge symmetries are not obstructed at one loop order. Otherwise, the theory presents an anomaly

\begin{equation}
\label{23}
{\cal A }[\,\phi, \phi^\ast \,]\, = \, \Delta S + { i }
( S , M_1 ) \,=\,a_\alpha\,c^\alpha+\dots\,.
\end{equation}

\bigskip
The nilpotency of the BRST operator implies that 
$s{\cal A}=0$, which is the Wess-Zumino consistence condition.
So, looking for possible anomalies in any theory is the same as looking for local functionals with ghost number one that are BRST closed ($s{\cal A}=0$) but not BRST exact (${\cal A}\neq sB$).

\bigskip
By using cohomological arguments, we can show that  the quantum master
equation, for first order systems with pure second class constraints converted with the use of the BFFT procedure, can  always be solved. To prove this, let us first derive the BRST transformations of the fields and antifields for the converted system:

\begin{eqnarray}
\label{a2}
s\,y^\mu&=&\{y^\mu,\tilde\chi_\alpha\}c^\alpha\nonumber\\
s\,\phi^\beta&=&\{\phi^\beta,\tilde\chi_\alpha\}c^\alpha\nonumber\\
s\,\lambda^\alpha&=&\dot c^\alpha\nonumber\\
s\,c^\alpha&=&0\nonumber\\
s\,\bar c_\alpha&=&\bar\pi_\alpha\nonumber\\
s\bar\pi_\alpha&=&0\nonumber\\
s\, y_\mu^*&=&-{{\partial S}\over{\partial y^\mu}}    \nonumber\\
s\,\phi_\alpha^*&=& -{{\partial S}\over{\partial \phi^\alpha}}   \nonumber\\
s\,\lambda_\alpha^*&=& \tilde\chi_\alpha   \nonumber\\
s\,c_\alpha^*&=& -y^*_\mu\{y^\mu,\tilde\chi_\alpha\}
-\phi^*_\beta\{\phi^\beta,\tilde\chi_\alpha\}-\dot\lambda^*
   \nonumber\\
s\,{\bar c}_\alpha^*&=& 0   \nonumber\\
s\,{\bar \pi}^{*\alpha}&=& {\bar c}^{*\alpha}  
\end{eqnarray}
\bigskip

\noindent where $S$ is given by (\ref{15}).
We see that ${\bar c_\alpha}$ and $\bar\pi_\alpha$ form
BRST doublets ($s\,B=C\,,s\,C=0$) and do not belong to the BRST cohomology \cite{HT}. The same is true for
their antifields. To show that  the other fields and antifields do not contribute to the cohomology  at ghost number one, it is enough to study the
cohomology   of the linearized piece of $s$, which will be denoted by $s^{(1)}$\cite{s1}. If we assume that  in the process of conversion of the constraints (see Eq. (\ref{9})), the invertible matrix $X(y)$  can be written as a power series in $y$ (which will be the case for the example we are going to consider),

\begin{equation}
\label{a3}
X(y)_{\alpha\beta}=X^{(0)}_{\alpha\beta}+
X^{(1)}_{\alpha\beta\mu}y^\mu+
X^{(2)}_{\alpha\beta\mu\nu}y^\mu y^\nu+\dots
\end{equation}

\noindent we see that

\begin{eqnarray}
\label{a5}
s^{(1)}\,\phi^\alpha&=&\omega^{\alpha\gamma}X^{(0)}_{\beta\gamma}c^\beta
\nonumber\\
s^{(1)}\,c^\alpha&=&0\nonumber\\
\end{eqnarray}

\noindent The equations above imply that 
$\phi^\alpha$ and $C^\alpha=\omega^{\alpha\gamma}X^{(0)}_{\beta\gamma}c^\beta$ form doublets under the action of $s^{(1)}$ and as a consequence they also do not belong to the cohomology . As $c^\alpha$ is trivially obtained from $C^\alpha$, and since it is the only fundamental field (or antifield) with positive ghost number,
it is not possible to construct a local functional with ghost number one that is BRST closed  not being BRST exact. This means that any candidate to an anomaly can always be canceled by some counterterm $M$. So the situations found in \cite{dWG} and later explored in \cite{us} appear also here:
enlarged symmetries due to compensating fields ( here the BFFT variables )
are not anomalous . This does not  mean that they have a trivial role at the quantum level  since the existence of a counterterm modify expectation values of relevant physical quantities \cite{us}. In the next section we are going to show an
example where all of these features
are carefully taken in account in order to derive  consistent quantum actions.

\bigskip
\section{
Massive  vector fields coupled to chiral\\
fermions                              }
\setcounter{equation}{0}
\bigskip

We shall now apply the ideas discussed above to massive chiral electrodynamics. Although the fermions couple only one chirality to the connection $A_\mu$, the second class system presents no gauge anomaly since it exhibits no gauge symmetry. When it is converted to a first class one, however, the fermions pass to transform in a chiral way  and such a gauge transformation is known to lead to possible anomalies \cite{ABJ}. Accordingly to the ideas discussed in the last section, however, the BFFT variables play the
role of Wess-Zumino fields and
permit us to write the anomaly candidates as BRST exact functionals,
solving in this way the quantum master equation at one loop order.

\bigskip
We start by considering the first order action

\begin{equation}
 S_0= \int d^4x \left\{ \dot A_\mu\pi^\mu
                      +i\bar\psi\gamma^0\dot\psi
                      -{\cal H}
                      -\lambda^\alpha\chi_\alpha
              \right\}
\label{y1}
\end{equation}

\noindent where the second class constraints

\begin{eqnarray}
\chi_1&=&\pi^0\nonumber\\
\chi_2&=&\partial_i\pi^i-m^2A^0+J^0
\label{y2}
\end{eqnarray}

\noindent and the first class Hamiltonian

\begin{eqnarray}
\label{y3}
H&=& \int d^{3}x \left\{
\frac{1}{2}\pi_i^2+\frac{1}{4}F_{ij}^2+\frac{1}{2}m^2\left(A_0^2+A_i^2
\right)
\nonumber\right.\\&&\left.
-i\bar\psi\gamma^iD^+_i\psi+\partial_i A^i \chi_1-A_0\chi_2 
\right\}
\end{eqnarray}\noindent have been introduced. In the above expressions
we have defined the covariant derivatives $D_\mu^+$ acting on the fermion  $\psi$  and the chiral projectors $P^\pm$ respectively as

\begin{eqnarray}
\label{y4}
D_\mu^+&=&\partial_\mu-ieP^+A_\mu\nonumber\\
P^{\pm}&=&\frac{1}{2}\left(1\pm\gamma^5\right)
\end{eqnarray}
We  have also adopted the metric convention $\eta^{\mu\nu}=\mbox{diag}(-1,+1,+1,+1)$. Dirac matrices satisfy the usual anticommutation relation  $\{\gamma^\mu,\gamma^\nu\}=2\eta^{\mu\nu}$. As one can verify, action
(\ref{y1}) is the first order version of

\begin{equation}
\label{y5}
{{\cal S}_{cov}}=\int d^4x\left[-\frac{1}{4}F_{\mu\nu}F^{\mu\nu}
           -\frac{1}{2}m^2A_\mu A^\mu
           +i\bar\psi\gamma^\mu D^+_\mu\psi\right]
\end{equation}

From (\ref{y1}) we extract the fundamental (equal time) brackets

\begin{equation}
\label{y6}
\left\{\psi(x),\bar\psi(y)\right\}=i\gamma^0\delta^3(x-y)
\end{equation}

\noindent for the fermionic sector and

\begin{equation}
\label{y7}
\left\{A_\mu(x),\pi^\nu(y)\right\}=\delta^\nu_\mu\delta^3(x-y)
\end{equation}

\noindent for the bosonic one. By using the above expressions, one can show, for instance, that the fermionic chiral current

\begin{equation}
\label{y8}
J^\mu\equiv\bar\psi\gamma^\mu P^+\psi
\end{equation}
 has brackets between its components given by
\begin{eqnarray}
\label{y9}
\left\{J^\mu(x),J^\nu(y)\right\}&=&ie^2\bar\psi M^{\mu\nu}P^+\psi\,\delta^3(x-y)\nonumber\\
M^{\mu\nu}&=&\gamma^\mu\gamma^0\gamma^\nu- \gamma^\nu\gamma^0\gamma^\mu                                       
\end{eqnarray}

 It is now easy to verify that the constraints and the Hamiltonian satisfy the bracket structure

\begin{eqnarray}
\{ \chi_1(x),\chi_2(y) \}&=& -m^2\delta^3(x-y)\nonumber\\
\{ \chi_1(x),H\}&=&\chi_2(x)\nonumber\\
\{ \chi_2(x),H\}&=&\partial_i\partial^i\chi_1(x)
\label{y10}
\end{eqnarray}

Let us now use the BFFT algorithm for implementing the Abelian conversion of the above bracket structure. As we have two second class constraints,
we introduce two BFFT variables $\phi^\alpha$, $\alpha=1,2$, and 
for simplicity demand that they
satisfy
\begin{equation}
\label{51}
\{ \phi^\alpha(x),\phi^\beta(y)\}=\epsilon^{\alpha\beta}\delta^3 (x-y)
\end{equation}
which gives the matrix $\omega^{\alpha\beta}$ as in Eq. (\ref{6}). In (\ref{51}) $\epsilon^{12}=-\epsilon^{21}=1$, $\epsilon^{11}=
\epsilon^{22}=0$. A possible solution to Eqs. (\ref{8}) via (\ref{9}-\ref{10a}) is achieved with \cite{BFFT}
\begin{eqnarray}\label{53}
{\tilde\chi}_1&=&\chi_1-m^2\phi^2\nonumber\\
{\tilde\chi}_2&=&\chi_2+\phi^1\nonumber\\
{\tilde H}&=&H+\int d^3x \left[
                                   \frac{1}{2m^2}(\phi^1)^2
                                   +\frac{1}{2}m^2{(\partial_i\phi^2)}^2
                                   -\frac{\phi^1}{m^2}\tilde\chi_2
                                   -\phi^2\nabla^2\tilde\chi_1
                             \right]\nonumber\\
&=& \int d^3x \left[\frac{1}{2}\pi^2_i+\frac{1}{4}F^2_{ij}
                   +\frac{1}{2}m^2\left(\tilde A_0^2+\tilde A_i^2\right)
                   -i\bar\psi\gamma^i D^+_i\psi
                   \right.\nonumber\\&&\left.
                   -\tilde A_0\tilde\chi_2
                   +(\partial_i\tilde A^i)\tilde\chi_1 \right]
\end{eqnarray}
where we have defined the quantities
\begin{eqnarray}
\tilde A_i&=&A_i-\partial_i\phi^2\nonumber\\
\tilde A_0&=&A_0+{\phi^1\over{m^2}}
\label{tilde}
\end{eqnarray}
Correspondingly we have a first order action
\begin{eqnarray}
 S_0&=& \int d^4x\left\{ \dot A_\mu\pi^\mu
                      +\dot\phi^1\phi^2
                      +i\bar\psi\gamma^0\dot\psi
                      -\tilde{\cal H}
                      -\lambda^\alpha\tilde\chi_\alpha
              \right\}
\label{54}
\end{eqnarray}
which is invariant under the gauge transformations generated by $\tilde\chi_1$ and 
$\tilde\chi_2$ (see Eq. (\ref{12}))
\begin{equation}
\begin{array}{ll}
\delta\psi=-ie\epsilon^2P^+\psi\,\,\,\,\,\,\,\,\,\,\,\,\, &
\delta\bar\psi=ie\epsilon^2\bar\psi P^- \\
\delta A_0=\epsilon^1 &
\delta\pi^0=-m^2\epsilon^2 \\
\delta A_i=-\partial_i\epsilon^2     &
\delta\pi^i=0         \\
\delta\phi^1=-m^2\epsilon^1     &
\delta\phi^2=-\epsilon^2    \\
\delta\lambda^1=\dot\epsilon^1      &
\delta\lambda^2=\dot\epsilon^2
\label{55}
\end{array}
\end{equation}

In the expressions above $\epsilon^\alpha$ are arbitrary space-time dependent parameters. We note that the variables $\tilde A_\mu$ are
invariant under (\ref{55}).
\bigskip

In order to quantize this system along the lines of the field-antifield formalism, 
associated with the parameters $\epsilon^\alpha$  we introduce the ghosts 
$c^\alpha$. We introduce also the trivial pairs $\bar\pi_\alpha$, $\bar c_\alpha$ and write down a gauge-fixed vacuum functional as in (\ref{19})
with 

\begin{eqnarray}
\label{66}
S_{nm}&=&S_0+\int dx^D \left[ A^{0*}c^1-m^2\pi_0^*c^2-A^{i*}\partial_ic^2 
\right.\nonumber\\&&\left.
-m^2\phi_{1}^*c^1-\phi_{2}^*c^2 +\lambda_1^*\dot c^1+\lambda_2^*\dot c^2
\right.\nonumber\\&&\left.
-ie\psi^*P^+\psi c^2+ie\bar\psi P^-\bar\psi^* c^2
+\bar\pi_\alpha \bar c^{\alpha*}
\right]
\end{eqnarray}
where some proper gauge-fixing fermion $\Psi$ is assumed.
Now  observe that the terms in $S_{nm}$ which involve the matter fields are
\begin{equation}
\label{68}
i\bar\psi\left[
\gamma^0\left(\partial_0-ieP^+(\tilde A_0-\lambda^2)\right)+\gamma^iD_i^+
\right]\psi
\end{equation}
The quantities $\bar A_0=\tilde A_0-\lambda^2$ and $\bar A_i=A_i$ transform
as $s\bar A_\mu=-\partial_\mu c^2$. As the fermions also transform consistently, as can be seen from (\ref{55}), we  obtain the action of
the operator $\Delta$ over $S_\Psi$ adopting canonical procedures. For instance, in a Pauli-Villars regularization scheme with a fermionic mass term
with usual form, which means that the vector symmetry is taken as a preferential one, we see that

\bigskip
\begin{equation}
\label{58}
\Delta S_\Psi=-{1\over {96\pi}}\int d^4x c^2\epsilon^{\mu\nu\rho\sigma}\bar F_{\mu\nu}\bar F_{\rho\sigma}
\end{equation}

\bigskip
\noindent where $\bar F_{\mu\nu}=\partial_\mu\bar A_\nu-\partial_\nu\bar A_\mu$ and possible normal parity terms in the original space of fields have been discarded. Eq. (\ref{58}) represents the essential candidate to the anomaly.
It is easy to see, however, that 

\begin{equation}
M_1={i\over{96\pi}}\int d^4x\phi^2\epsilon^{\mu\nu\rho\sigma}\bar F_{\mu\nu}\bar F_{\rho\sigma}
\label{M1abel}
\end{equation}
 solves the one loop master equation, which means that
we have achieved a consistent route for the quantization of the theory. The gauge fixed vacuum functional reads
\begin{equation}
\label{Z-abel}
Z=\int [d\Phi^A]
\exp\left\{\frac{i}{\hbar} W [\Phi^A,\Phi^*_A=\frac{\partial\Psi}{\partial\Phi^A}] \right\}
\end{equation}
with $ [d\Phi^A]=(A_\mu,\pi^\mu,\phi^\alpha,\psi,\bar\psi,\lambda^\alpha,c^\alpha,
\bar c_\alpha,\bar\pi_\alpha) $, and all possible information about the system can be obtained from it.
If we wish to write an effective quantum action in an explicitly covariant way we may eliminate the momenta through functional
integrations in (\ref{Z-abel}). Let us assume that the gauge fixing fermion $\Psi$ does not depend on $\lambda^1$ or $\pi^\mu$, consequently $\lambda_1^*=\pi^*_\mu=0$. Suppose also that $\Psi$ possibly depends on $\lambda^2$ only through an $\bar A_0$ dependence. Integration in $\lambda^1$ and $\pi^0$ results in the substitution
$
\pi^0
\rightarrow
m^2 \phi^2
$ in $W$. Under the redefinition
\begin{equation}
A_0\longrightarrow A_0+\lambda^2-\frac{\phi^1}{m^2}
\label{cvabel}
\end{equation}
we obtain the intermediate auxiliary quantum action
\begin{eqnarray}
W_{aux}&=&\int d^4x \left[(A_0+\lambda^2)\dot\phi
+\dot A_i\pi^i
+i\bar\psi\gamma^0\dot\psi
-\frac{1}{4}F_{ij}^2
-\frac{1}{2}{\pi^i}^2
\nonumber\right.\\&&\left.
-\frac{1}{2}m^2(A_0+\lambda_2)^2
-\frac{1}{2}m^2\left(A_i-{\partial_i\phi}\right)^2
+i\bar\psi\gamma^i D_i\psi
\nonumber\right.\\&&\left.
+A_0\left(\partial_i\pi^i+J^0+m^2(A_0+\lambda^2)\right)\right]
+M_1+S_{\mbox{gf}}
\label{Waux}
\end{eqnarray}
where
\begin{eqnarray}
S_{\mbox{gf}}&=&\int d^4x \left[ 
-\frac{\delta\Psi}{\delta A_\mu}\partial_\mu c^2
-m^2\frac{\delta\Psi}{\delta \phi^{1}} c^1
-\frac{\delta\Psi}{\delta\phi^2}c^2
\right.\nonumber\\&&\left.
-ie\frac{\delta\Psi}{\delta\psi}P^+\psi c^2
+ie\bar\psi P^-\frac{\delta\Psi}{\delta\psi}c^2+\bar\pi_\alpha
\frac{\delta\Psi}{\bar c_\alpha}
\right]
\end{eqnarray}
and $M_1$  is given by (\ref{M1abel}) without the bars in $F_{\mu\nu}$ because of (\ref{cvabel}).
Further integration in $\lambda_2$ and $\pi^i$ results in the effective 
quantum
action\bigskip
\begin{equation}
W_{\mbox{eff}}=\int d^4x \left[
               -\frac{1}{4}F^2_{\mu\nu}
               -\frac{1}{2}m^2{\left(A_\mu-{\partial_\mu\phi^2}\right)}^2
               +i\bar\psi\gamma^\mu D^+_\mu \psi
\right]
               +M_1
               +S_{\mbox{gf}}
\end{equation}\bigskip

As we have already mentioned, a convenient choice of $\Psi$ fixes all the
gauge symmetry of the theory. We cite some possible choices for $\Psi$. The unitary gauge is achieved with $\Psi=\int d^4x\bar c_\alpha \phi^\alpha$ followed by functional integration on $\bar\pi_\alpha$ and $\phi^\alpha$.  With this choice the quantum action  reduces to the simple form 
(\ref{y5}) and the path integral presents the usual Lioville's measure
for the pertinent fields.
The choice
$\Psi=\int d^4x\left[\bar c_2 ({{\alpha\bar\pi^2}\over{2}}+\partial_\mu A^\mu)+\bar c_1\phi^1\right]$
leads to the usual covariant Gaussian gauge fixing depending on the arbitrary parameter $\alpha$. In this situation 
\bigskip

\begin{eqnarray}
S_{\mbox{gf}}&=&\int d^4x \left[ 
-\partial^\mu\bar c_2\partial_\mu c^2
+\bar\pi_2({{\alpha\bar\pi^2}\over2}+\partial_\mu A^\mu)
+\bar\pi^1\phi^1-m^2\bar c_1 c^1
\right]
\end{eqnarray}
and the integration over $c^1,\,\bar c_1,\,\bar \pi_1,\,\phi^1$ is trivial. 

An interesting situation comes if we fix the compensating field $\phi^2$ to
some external value, say, $\phi^2=\beta$. By choosing $\Psi=\bar c_1\phi^1+ \bar c_2(\phi^2-\beta)$, we obtain, after a few trivial integrations and the absorption of some trivial normalization factors by the measure, that

\begin{equation}
Z[\beta]=\int [d\psi][d\bar\psi][d A^\mu]\exp\left\{{i\over\hbar} W_{ext}[\psi,\bar\psi,A,\beta]\right\}
\label{zbeta}
\end{equation}
where
\begin{eqnarray}
\label{Wbeta}
W_{ext}[\psi,\bar\psi,A,\beta]&=&\int d^4x [
               -\frac{1}{4}F^{\mu\nu}F_{\mu\nu}
               -\frac{1}{2}m^2\left(A_\mu-\partial_\mu\beta\right)^2\nonumber\\
              &+&i\bar\psi\gamma^\mu D^+_\mu \psi
+{{i\hbar}\over{96\pi}}\beta\epsilon^{\mu\nu\rho\sigma}
 F_{\mu\nu} F_{\rho\sigma}]
\end{eqnarray}

The condition that the path integral cannot depend on $\beta$, which comes from the 
Fradkin-Vilkoviski theorem, gives, for instance,
that 

\begin{equation}
\label{expectation}
i\hbar{{\delta Z[\beta]}\over{\delta\beta}}{\mid_{\beta=0}}
=<m^2\partial_\mu A^\mu + {{i\hbar}\over{96\pi}}\epsilon^{\mu\nu\rho\sigma}
 F_{\mu\nu} F_{\rho\sigma}>_{\mid_{\beta=0}}=0
\end{equation}
which is a surprising result. If we observe, however, that
$\partial_\mu J^\mu=-m^2\partial_\mu A^\mu$ as a consequence of the equations of motion for the field $A_\mu$ in the unitary gauge,
we can interpret Eq. (\ref{expectation})
as the anomalous divergence of the Noether current (\ref{y8})
associated to the rigid chiral symmetry present in the original
theory given by actions (\ref{y1}-\ref{y5}). This is an unexpected result
derived from the quantum BFFT formalism. Similar results have recently been derived by using compensating fields at Lagrangian level \cite{us}. In these
last approaches, the compensating fields coupled directly to the chiral
current in an extended $QCD$ which presents not only vector but also
chiral gauge symmetry.

\section{Conclusions}

In this work we have considered the BFFT quantization of first order systems submitted to pure second class constraints. We have shown that the gauge symmetries introduced by the BFFT procedure are not obstructed at quantum level, since the compensating fields do not belong to the BRST cohomology at ghost number one. An specific example has been given, where massive electrodynamics couples to chiral fermions. The quantum master equation has been 
solved and the corresponding counterterm has played an essential role in extracting anomalous expectation values of physically relevant quantities. 
We would like to finish by commenting that a few generalizations could have
been considered. We could have started from an already gauge invariant first order system with both first and second class constraints. Then it would be necessary to take care of both symmetry sectors, the original one and that introduced by the BFFT conversion procedure.  Another
possibility could be considering examples with more involving algebraic structure, as it occurs with some of the models cited in Ref. \cite{BFFT}.
We are now studying aspects of these subjects and results will be reported  elsewhere \cite{AT}.

\vskip 1cm
\noindent {\bf Acknowledgment:} We are in debt to 
N. R. F. Braga for an useful discussion. 
This work is supported in part by
Conselho Nacional de Desenvolvimento Cient\'{\i}fico e Tecnol\'ogico
- CNPq (Brazilian Research Agency).

\end{document}